\documentclass[aps,pra,twocolumn,showpacs,preprintnumbers]{revtex4}

\usepackage{mathbbol}
\usepackage{bbm}
\usepackage{graphicx}

\newcommand{\tr}[1]{\ensuremath{ \mathrm{tr}\left( #1 \right) }}
\newcommand{\trz}[2]{\mathrm{tr}_{#1}\! \left( #2 \right)}

\newcommand{\etal}{{\it et al.\ }}

\newcommand{\bra}[1]{\left< #1\right|}
\newcommand{\ket}[1]{\left| #1\right>}

\newcommand{\expect}[1]{\left< #1 \right>}
\newcommand{\abs}[1]{\left| #1 \right|}
\newcommand{\1}{\mathbbm{1}}
\newcommand{\proj}[1]{\ket{#1}\bra{#1}}

\begin{document}

\title{Entanglement witnesses and a loophole problem}
\author{Patrick Skwara}
\author{Hermann Kampermann}
\author{Matthias Kleinmann}
\author{Dagmar Bru\ss }
\email{bruss@thphy.uni-duesseldorf.de}
\affiliation{Institut f\"ur Theoretische Physik III, Heinrich-Heine-Universi\"at D\"usseldorf, D-40225
D\"usseldorf, Germany}


\date{\today}

\begin{abstract}

We consider
a possible detector-efficiency loophole
in experiments that detect entanglement via the
local measurement of
witness operators.  Here, only local properties of
the detectors are known.
We derive a general threshold for the detector
efficiencies which guarantees that a negative
expectation value of a witness is due to entanglement,
rather than to erroneous detectors.
This threshold depends on the local decomposition
 of the
witness and its  measured expectation value.
For two-qubit witnesses
we find the  local operator decomposition that is optimal
with respect to closing the loophole.

\end{abstract}

\maketitle

\section{Introduction}
Entanglement is the central theme in quantum information processing.
It allows to design faster algorithms than classically,
 to communicate in a secure way, or to perform protocols that
have no classical analogue.
Entangled states of few particles (e.g. photons, ions) can be routinely
created in experiments, and their entanglement can be confirmed using
state tomography, Bell inequalities or entanglement witnesses.
All of these tools are well-established methods for the detection of
entanglement. But can one be sure that they give a confirmative answer
even when realistic, i.e. erroneous detectors are used? Here,
we will introduce
and study a loophole-problem for the detection of entanglement
via witness operators.

Loophole-problems have been widely discussed
in the context of ruling out
local hidden variable (LHV) models,
by measuring
a violation of certain inequalities, as suggested in the seminal work of
J.S. Bell in 1964
\cite{Bell}.
Many experiments have been carried out along
 that line \cite{Aspect1},
but  all of them  so far suffer from
 the  locality
loophole (i.e. no causal separation of the detectors) and/or
the  detection  loophole (i.e. low detector efficiency).
As a consequence of a loophole,  quantum correlations are also explainable by
LHV theories
\cite{BellUnSpeak,Larsson1}.

In this paper we discuss
 a possible  {\em detection loophole} for experiments
that measure entanglement witnesses.
Here, the
goal is not to prove the completeness of quantum mechanics
(as in Bell experiments),
but, assuming the correctness of quantum mechanics,
to prove the existence of entanglement in a given state.
One advantage of witness operators is that they require only
few {\em local} measurements to detect entanglement \cite{Guehne};
 global measurements are experimentally not easily
accessible at present. A local
projection
measurement with realistic imperfect detectors
(in the computational basis, for qubits and isotropic noise)
 can be described by the
following  positive operator valued measurement (POVM):
$F_0 = \xi \proj{0}, F_1 = \xi \proj{1},$ and
$ F_2 = (1-\xi)\1$,
where $\xi$ is the efficiency of the detector.
 However, in general the {\em global} properties of the detectors
are not fully characterised, e.g. there may exist correlations between POVM
elements of different detectors.  Provided that only local detector
properties are given, what are the conditions for being nevertheless
able to prove the existence of entanglement without doubt?

An entanglement witness $W$
is a Hermitian operator that
fulfils
$\tr{W\rho_s}\geq 0$ for all separable
$n$-partite states
$\rho_s=\sum_ip_i\bigotimes_\nu^n\big|\psi_{\nu}^{(i)}\big>\big<\psi_{\nu}^{(i)}\big|$  \cite{Werner},
where the index $\nu$ numbers the subsystem, the probabilities
$p_i$ are real and non-negative, $\sum_ip_i=1$,
and $\tr{W\rho_e}<0$ for at least one entangled state $\rho_e$
\cite{WitnessH,Lewenstein}.
Throughout this paper, we will use without loss of generality normalized witnesses,
i.e. $\tr{W}=1$.
Our goal is to ensure that a negative measured expectation value $\expect{W}_m<0$
is really due to  the state being entangled, rather than to
  imperfect detectors.
The following line of arguments  also holds for  specialized
witnesses which are constructed such that, e.g.,
 they detect only genuine multi-partite entanglement
\cite{BrussW1}
or states prohibiting LHV models \cite{Hyllus}.

This paper is organized as follows. After introducing the
local decomposition of entanglement witnesses, we study the effect
of lost events as well as additional events on
the experimental expectation value of $W$. Here, we use the worst case
approach  to derive inequalities which need to be
fulfilled to ensure entanglement of the given state. The parameters in
these inequalities are the measured expectation value of the witness, the
detection efficiencies, and the coefficients for the local decomposition
of the witness.
We show for the two-qubit case how to optimize the local
operator decomposition of the witness,
such that the
 detection efficiency
which is needed to close the loophole is minimized. Some recent
experiments measuring witness operators
are presented,
to demonstrate that the detection loophole is
 a problem in current experiments.

\section{Detection loophole for entanglement witnesses}
Any witness for an $n$-partite quantum state in $d=\Pi_\nu^n d_\nu$ dimensions
 can be decomposed in a local operator basis, i.e.
an $n$-fold tensor product of  operators $\sigma_{\nu}^{(i)}$,
\begin{equation}\label{Eq-WDec}
W=\sum_{i =0} c_i\bigotimes_\nu^n \sigma_{\nu}^{(i)},
\label{witness}
\end{equation}
where the coefficients $c_i$ are real.
Each  operator $\sigma_{\nu}^{(i)}$ is traceless or the identity and
corresponds to the
$i$th local setting for the party number $\nu$.
In this expansion, we include implicitly also the local identity operators which do not need
to be measured. The number of terms in eq.\ (\ref{Eq-WDec}) depends on the decomposition,
i.e.\ on the choice of operators $\sigma_{\nu}^{(i)}$.
 A straightforward, but not necessarily optimal choice (concerning the
needed detector efficiency) are the
$d_\nu^2-1$ Hermitian
generators $\sigma_{\nu}$ of SU($d_\nu$) and the respective identity operators.

In the following, we will use a simpler notation, namely
\begin{equation}
W=c_0\mathbbm{1}+\sum_{\alpha=1} c_\alpha S_\alpha,
\label{witnesssimple}
\end{equation}
where
$S_\alpha$ stands for one term
from the local expansion (\ref{witness}).
Here,  we exclude  the identity $\mathbbm{1}$
(acting on the total space)
 from the sum over $\alpha$, because
it does not have to be measured and therefore has a special role.

We will now investigate
 one local measurement setting described by $c_\alpha S_\alpha$,
and drop the index $\alpha$ for convenience.
The \emph{ measured} expectation value $c\expect{S}_m$ is given by
\begin{equation}
\label{Eq-SettingExpect}
c\expect{S}_m=c\, \frac{\sum_i n_i \lambda_i}{N}=
c\, \frac{\sum_i \left(\tilde{n}_i+\varepsilon_{+i}-\varepsilon_{-i}\right)
\lambda_i}{\tilde N+\varepsilon_{+}-\varepsilon_{-}},
\end{equation}
where $\lambda_i$ is the $i$th eigenvalue of $S$,
the number of measured events for the
$i$th  outcome  is denoted as $n_i$, and $N$ is the total number of measured
events for that setting.
In the second part
of eq. (\ref{Eq-SettingExpect}) we expressed this expectation
 value as a sum of the ideal number of events, denoted as  $\tilde{n}_i$ (i.e.
for perfect detectors),
the additional events $\varepsilon_{+ i}$ (e.g.\ dark counts)
and the lost events $\varepsilon_{- i}$ 
for the $i$th outcome. The total ideal number of events for such a
 setting
is denoted as $\tilde{N}$, and the total number of additional/lost
events as $\varepsilon_\pm$.
We have $\tilde{N}=\sum_i \tilde{n}_i$
and $\varepsilon_\pm=\sum_i\varepsilon_{\pm i} $.

The experimental data usually gives no information
about the  number of errors $\varepsilon_{\pm i} $
for a specific measurement outcome $i$. Only
the detection imperfections
are known,
namely  the detection efficiency (``lost events efficiency''):
\begin{equation}
\eta_-=\frac{\tilde{N}-\varepsilon_-}{\tilde{N}}
\end{equation}
and the ``additional events efficiency'':
\begin{equation}
\eta_+=\frac{\tilde{N}}{\tilde{N}+\varepsilon_+},
\end{equation}
where $0\leq \eta_{\pm} \leq 1$ holds.
Here, $\eta_{\pm}$ denotes the  global detection efficiency
for a given  measurement setting. In the following, we assume
that $\eta_{\pm}$ is the same for all settings. (Other cases
can be included by indexing
$\eta_{\pm}$ with $\alpha$.)

Usually, one makes the fair-sampling assumption about the
statistical distribution of the unknown errors
$\varepsilon_{\pm i}$,
 i.e.\ one assumes the same
statistical distribution for detected and lost events;
the additional events are assumed to have a flat distribution.
Here,  we will give up this  assumption and
 will consider the worst case, where both lost and additional
events contribute such that the expectation value of the witness
is shifted towards negative values.
We point out that in order
to reach this worst case scenario,
it is already sufficient
that the global POVM elements  exhibit certain classical correlations
(while being compatible with the
local measurement operators)
\cite{tobepu}.

The worst case is equivalent to finding the
lowest possible $c\expect{S}_m$, which is achieved by
minimizing the contribution of the additional events, and maximizing
the contribution of the lost events in
eq.\ (\ref{Eq-SettingExpect}).
 This minimization/maximization can be easily shown to have the form
$\min_{\varepsilon_{+i}}c\sum_i \varepsilon_{+i}\lambda_i=\varepsilon_{+}\Omega_{+}$ and
$\max_{\varepsilon_{-i}}c\sum_i \varepsilon_{-i}\lambda_i=\varepsilon_{-}\Omega_{-}$,
with
\begin{equation}
\Omega_{\pm}=c\left(\Theta(\pm c)\lambda_{\mathrm{min}}+\Theta(\mp c)\lambda_{\mathrm{max}}\right),
\end{equation}
where $\Theta(x)$ denotes the Heaviside function, and
$\lambda_{\mathrm{min/max}}$ is the minimal/maximal eigenvalue of $S$.

Inserting this into eq.\ (\ref{Eq-SettingExpect}), one finds the following
worst case estimate for the measured expectation value $c\expect{S}_m$:
\begin{equation}
c\expect{S}_m
 =  \kappa\left(c\expect{S}_t-\Omega_+(1-\frac{1}{\eta_+})-\Omega_-\left(1-\eta_-\right)\right),
\label{expectvalue}
\end{equation}
where $\kappa:=\tilde N \left(\tilde N +\varepsilon_+-\varepsilon_-\right)^{-1}=
\left(\frac{1}{\eta_+}+\eta_--1\right)^{-1}$.
 Here, we have introduced the notation
$c\expect{S}_t$ for the \textit{true} expectation value (without any errors).

Using $\expect{W}_{m/t}=c_0+\sum_\alpha c_\alpha\expect{S_\alpha}_{m/t}$
and eq. (\ref{expectvalue}), with isotropic detection efficiencies, we can
express the  measured witness expectation value
$\expect{W}_m$ as a function  of the \emph{true} one $\expect{W}_t$,
for the worst case.
To close the loophole it is necessary to ensure that $\expect{W}_t<0$.
 This leads to a condition for  the
maximal $\expect{W}_m$ that depends on the decomposition of $W$
and the efficiencies $\eta_\pm$:
\begin{eqnarray}\label{Eq-WmIneq}
\expect{W}_m & < & c_0(1-\kappa)-
\nonumber \\
& & \kappa \sum_\alpha\left(\Omega_{\alpha +}\left(1-\frac{1}{\eta_+}\right)
+\Omega_{\alpha -}\left(1-\eta_-\right)\right)\! \! , \:{ }
\end{eqnarray}
where we have re-introduced the summation index $\alpha$.

\begin{figure}
\scalebox{0.6}{\includegraphics{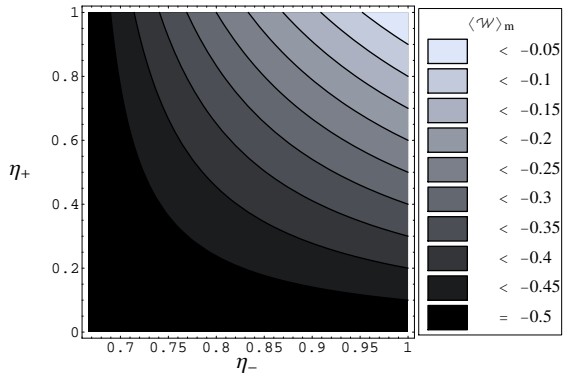}}
\caption{\label{Fig-WmEtpEtm}
The contours correspond to the maximal $\expect{W}_m$ of a multipartite qubit witness which still ensures entanglement,
where a decomposition with $c_0+\sum_\alpha \abs{c_\alpha}=1$ was assumed, which is e.g.\ achieved for an
optimal two qubit Bell state witness $W_{\phi^+}$ (see text).
}
\end{figure}

>From now on we want to focus on the case where the subsystems are
 two-dimensional,
i.e. qubits. For qubits the
measurement operators $S_\alpha$ are chosen to be tensor products of Pauli operators with
eigenvalues $\lambda=\pm 1$. This simplifies $\Omega_{\pm}$ to
$\Omega_{\alpha \pm}=\mp \abs{c_\alpha}$, and eq.\ (\ref{Eq-WmIneq})
reads for qubits
\begin{equation}\label{Eq-WmEtpEtmQbts}
\expect{W}_m  < c_0 -
\frac{c_0+\sum_\alpha \abs{c_\alpha}\left(\frac{1}{\eta_+}-\eta_-\right)}
{\eta_-+\frac{1}{\eta_+} -1} .
\end{equation}
In Fig.\ \ref{Fig-WmEtpEtm} a contour plot of this function is shown:
Given a certain measured expectation value $\expect{W}_m$, the corresponding
efficiencies $\eta_\pm$ ensure that the state is indeed entangled.
This plot assumes $c_0+\sum_\alpha\abs{c_\alpha}=1$,
and can be easily redrawn for other decompositions.

>From eq.\ (\ref{Eq-WmEtpEtmQbts})
 it is obvious that an \emph{ optimal} decomposition of such a witness
with respect to the needed efficiencies $\eta_\pm$ is
achieved by minimizing $\sum_\alpha \abs{c_\alpha}$.
For the case of two-qubit witnesses a constructive optimization can be achieved, when
arbitrary local Stern-Gerlach measurements (described by Pauli operators
or rotations thereof) and
the identity  are allowed.
We start with a two-qubit witness in its Pauli operator decomposition, i.e.\
\begin{eqnarray}
W & = & c_{00}\1\otimes\1 + \1\otimes\left(\sum_{j=1}^3 c_{0j}\sigma_{j}\right) + \nonumber\\
& & \left(\sum_{i=1}^3 c_{i0}\sigma_{i}\right)\otimes\1+
\sum_{i,j=1}^3 c_{ij}\sigma_{i}\otimes\sigma_{j},\label{Eq-2QWitnessTerms}
\end{eqnarray}
where $\sigma_{1/2/3}=\sigma_{x/y/z}$ and $c_{ij}\in \mathbbm{R}$.
The normalization condition $\trz{}{W}=1$ leads to $c_{00}=\frac{1}{4}$. The three remaining terms in eq.\
(\ref{Eq-2QWitnessTerms}) can be optimized separately (note the special role of the identity).

 Let us first
consider the term
$W_3=\sum_{i,j=1}^3 c_{ij}\sigma_i\otimes\sigma_j$.
This expression is optimized by doing a singular value decomposition of the
coefficient matrix $(C)_{ij}=c_{ij}$, i.e.\
$UCV^\dag=S$, where $S$ is the diagonal matrix that contains the singular
values $s_i$.
The matrices $U$ and $V$ are orthogonal and have entries $u_{ij}$ and $v_{ij}$.
The new orthogonal basis is simply constructed by using the orthonormal rows
of $U$ and $V$, i.e.\
$\tilde{\sigma}_{Ai}  =  \sum_j u_{ij}\sigma_{j}$ and
$\tilde{\sigma}_{Bi}  =  \sum_j v_{ij}\sigma_{j}$,
such that we get the Schmidt operator decomposition
\begin{eqnarray}\label{Eq-OptDec}
W_3 & = & \sum_{i=1}^3 s_i \tilde{\sigma}_{Ai}\otimes\tilde{\sigma}_{Bi},
\end{eqnarray}
with $\frac{1}{2}\trz{}{\tilde{\sigma}_{Ai}\tilde{\sigma}_{Aj}}=\delta_{ij}$ and the same orthogonality relation for party $B$.

The optimality of this biorthogonal decomposition with respect to the detector
efficiencies is shown as follows:
Consider the most general decomposition
$W_3=\sum_j^M b_j\sigma'_{Aj}\otimes\sigma'_{Bj}$,
where $\sigma'_{A/Bj}$ are arbitrary (not necessarily orthogonal) rotated Pauli operators
and without loss of generality $b_j>0$
(we can include a minus sign in one of the operators).
Here, the number of terms
 $M$ is finite and an operator may appear more than once.

We can express this decomposition in terms of our orthogonal basis $\tilde{\sigma}_{A/Bi}$,
\begin{equation}\label{Eq-GenDec}
W_3=\sum_j^M b_j\sigma'_{Aj}\otimes\sigma'_{Bj} =
\sum_{j,k,l} b_j\alpha_k^{(j)}\beta_l^{(j)}\tilde{\sigma}_{Ak}\otimes\tilde{\sigma}_{Bl},
\end{equation}
with $\sum_k\left(\alpha_k^{(j)}\right)^2=\sum_k\left(\beta_k^{(j)}\right)^2=1$ for all $j$.
The right hand sides of eq.\ (\ref{Eq-OptDec}) and eq.\ (\ref{Eq-GenDec}) are equal.
We multiply these two expressions by $\sum_m\tilde{\sigma}_{Am}\otimes\tilde{\sigma}_{Bm}$ and
take the trace on both sides. This leads to
\begin{eqnarray}
 \sum_{i,m} s_i \trz{}{\left[\tilde{\sigma}_{Ai}\otimes\tilde{\sigma}_{Bi}\right]\left[
\tilde{\sigma}_{Am}\otimes\tilde{\sigma}_{Bm}\right]}
 =  \nonumber\\
 \sum_{j,k,l,m} b_j\alpha_k^{(j)}\beta_l^{(j)} \trz{}{\left[\tilde{\sigma}_{Ak}\otimes\tilde{\sigma}_{Bl}
 \right]\left[
\tilde{\sigma}_{Am}\otimes\tilde{\sigma}_{Bm}\right]}.
\end{eqnarray}
Orthogonality of the basis is used to get
\begin{eqnarray}
  \sum_{i} s_i
&  = &   \sum_{j,m} b_j\alpha_m^{(j)}\beta_m^{(j)}
\leq  \sum_{j} b_j,
\end{eqnarray}
where we used the fact that the scalar product between two normalized vectors ($\vec \alpha^{\,(j)}$ and
$\vec \beta^{\,(j)}$) is less or equal to one. This proves the optimality of the decomposition of $W_3$
given in eq.\ (\ref{Eq-OptDec}).

The third term in eq.\ (\ref{Eq-2QWitnessTerms}) -- and, analogously, the second one -- can be written as
\begin{equation}\label{Eq-OptIdentPart}
W_2=\tilde{c}_{10}\left(\sum_{i=1}^3\alpha_i\sigma_i\right)\otimes\1=\tilde{c}_{10}\tilde{\sigma}_{10}\otimes\1,
\end{equation}
where
$\tilde{\sigma}_{10}$ is a rotated Pauli operator, $\sum_i\alpha_i^2=1$
and $\tilde{c}_{10}>0$. Following similar arguments as before, it is easy to verify that
$\tilde{c}_{10}$ is optimal.

The situation for higher dimensions is different:
It is unfortunately not straightforward to generalize the above
 optimization to higher-dimensional
witnesses, because we extensively used the fact that a linear combination of Pauli operators is again a
scaled rotated Pauli operator. Also, for multi-partite witnesses the Schmidt decomposition eq.\ (\ref{Eq-OptDec})
does not always exist, such that our optimization method
 is  not applicable
 for these cases.

In many experimental situations, e.g.\ when optical detectors are used,
only the ``lost event efficiency'' is an important issue
and the ``additional event efficiency'' is approximately $\eta_+\approx 1$ \cite{Barbieri,Bourennane}.
This situation further simplifies eq.\ (\ref{Eq-WmIneq}), and
 the minimal detector efficiency that allows to
 close the loophole has
 a simple relation
with  the measured expectation value and the decomposition of the witness, namely
\begin{equation}\label{Eq-EtaMWm}
\eta_->\left(1-\frac{\expect{W}_m}{c_0+\sum_\alpha \abs{c_\alpha}}\right)^{-1}.
\end{equation}
\begin{figure}
\scalebox{0.45}{\includegraphics{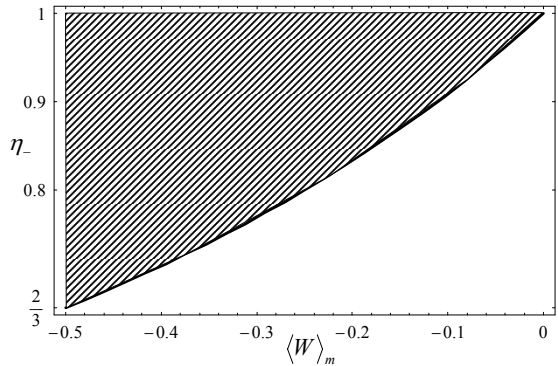}}
\caption{\label{Fig-EtaMWm}
The detector efficiency loophole is closed (assuming $\eta_+=1$) if the detector efficiency ($\eta_-$)
is in the hatched area for a given measured expectation value $\expect{W}_m$.
Here, a decomposition with $c_0+\sum_{\alpha} \abs{c_\alpha}=1$ is assumed.
}
\end{figure}
This function is shown in fig.\ (\ref{Fig-EtaMWm}), where
 the hatched area corresponds to values for which the
inequality is fulfilled, again
assuming $c_0+\sum_{\alpha} \abs{c_\alpha}=1$.
For two qubits the detection loophole for witnesses
can already be closed with a detection efficiency
$\eta_->\frac{2}{3}$.
This bound is sufficient e.g.\ for the optimal two-qubit witness of a Bell state
and a measured expectation value of $\expect{W}_m=-\frac{1}{2}$. We conjecture
that for $\eta_-<\frac{2}{3}$ the loophole cannot be closed for any witness.

\section{Experimental examples and summary}
We now want to discuss some prominent experimental examples in this context.
In ion trap experiments the detection efficiencies are close to one ($\sim 99.8\% $) \cite{ZollerIonTrap}
and the detection loophole is usually
not an issue. Exceptions are many-party-witnesses with expectation values close to zero, like the genuine 8-qubit multipartite entanglement
witness experiment of H\"affner \etal \cite{Haeffner}.

Single photon experiments on the other hand are more problematic. Using
eq.\ (\ref{Eq-EtaMWm}) we give some explicit experimental examples for the needed detection efficiencies
to close the witness loophole:
M. Barbieri \etal \cite{Barbieri}  implemented the optimal
two-qubit entanglement witness to detect a Bell state, where they
achieved an expectation value of $-0.493\pm 0.008$. In this case the detection efficiency needs to be $\eta_->0.67$.
In recent experiments also multipartite entanglement witnesses were implemented \cite{Bourennane}. In this work
the three-qubit GHZ entanglement witness
is loophole-free with a detection efficiency of $\eta_->0.91$, and the four-partite case needs $\eta_->0.94$.
Single photon detector efficiencies
for wavelengths of 700-800 nm
are typically around 70\%
\cite{DetectEffKwiat}, such that the global detection efficiency for two qubits
is circa 50 \% , and even lower for more than two subsystems.
The detection efficiencies for multipartite witness
experiments with photons are thus considerably  below the
needed thresholds. This is
a similar situation as for the detection loophole in Bell inequalities \cite{Larsson1}.
However, there is a good chance for loophole-free
witness experiments with two qubits, when slightly more efficient detectors
are available.

In summary, we discussed the detection loophole problem for
experiments measuring witness operators.
Assuming the worst case (that may occur  due to
 unknown global properties of the detectors)
we derived certain inequalities to
close such loopholes.
These inequalities  are generally valid for any type of witness operator
and
depend on the measured  expectation value of the witness,
its local operator decomposition and the detector efficiencies.
>From there,
 detector efficiency thresholds
to close the loophole  are easily calculated.
The local
decomposition of the witness  can be optimized such that
 the needed detection efficiencies are minimized.
We explicitly presented a constructive optimization for two-qubit witnesses.
For multi-qubit witnesses
the optimal decomposition is achieved by minimizing the sum of the
 absolute values of the expansion coefficients.
In the case of higher-dimensional witnesses the optimization
is not straightforward any more,
because it then also depends on the type of operator basis.
 Let us mention that an analogous study can be performed,
 if the witness is decomposed into local projectors
\cite{Guehne2}; this will be published elsewhere.
For qubit witnesses we further considered the common experimental
 situation, where additional counts can be neglected.
Current witness experiments with polarized photons do not close the detection
loophole, because of the low single photon detector efficiencies.
-- Further research directions and open problems include
the optimal local decomposition for higher-dimensional witnesses,
and the case of erroneous detector orientations.

We acknowledge discussions with
 Harald Weinfurter. This work
was supported in part by the European Commission (Integrated
Projects SCALA and SECOQC).


\begin{thebibliography}{Lorensen-87}
\addcontentsline{toc}{chapter}{Bibliography}
\label{Bibliography}
%
\bibitem{Bell}
    J. S. Bell, Physics (Long Island City, NY) {\bf 1}, 195 (1964)
\bibitem{Aspect1}
    A. Aspect, P. Grangier, and G. Roger, Phys. Rev. Lett. {\bf 47}, 460
(1981);
    A. Aspect, P. Grangier, and G. Roger, Phys. Rev. Lett. {\bf 49}, 91 (1982);
    A. Aspect, J. Dalibard, and G. Roger, Phys. Rev. Lett. {\bf 49}, 1804
(1982);
G. Weihs, T. Jennewein, C. Simon, H. Weinfurter, and A. Zeilinger,
 Phys. Rev. Lett. {\bf 81}, 5039 (1998);
    M.A. Rowe \emph{et al.}, Nature {\bf 409}, 791 (2001).
\bibitem{BellUnSpeak}
J. S. Bell, {\em Speakable And Unspeakable In
Quantum Mechanics} (Cambridge University Press), EPR-Experiments, (1987);
    T. K. Lo and A. Shimony, Phys. Rev. A {\bf 23}, 3003 (1981);
    A. Garg and N. D. Mermin, Phys. Rev. D {\bf 35}, 3831 (1987).
\bibitem{Larsson1}
    J.-\AA. Larsson, Phys. Rev. A {\bf 57}, 3304 (1998).
\bibitem{Guehne}
        O. G\"uhne, P. Hyllus, D. Bru\ss, A. Ekert,
   M. Lewenstein, C. Macchiavello,
and A. Sanpera, Phys. Rev. A
{\bf 66}, 62305 (2002).
\bibitem{Werner}
R.F. Werner, Phys. Rev. A, {\bf 40}, 4277 (1989).
\bibitem{WitnessH}
M. Horodecki \emph{et al.}, Phys. Lett. A, {\bf 223}, 1 (1996).
\bibitem{Lewenstein}
    M. Lewenstein, B. Kraus, J. I. Cirac, and P. Horodecki, Phys. Rev. A {\bf 62}, 052310 (2000).
\bibitem{BrussW1}
 D. Bru\ss \emph{ et al.}, J. Mod. Opt., {\bf 49}, 1399 (2002).
\bibitem{Hyllus}
        P. Hyllus, O. G\"uhne, D. Bru\ss, and M. Lewenstein,
Phys. Rev. A {\bf 72}, 12321 (2005).
\bibitem{tobepu} H. Kampermann \emph{et al.}, to be publ.
\bibitem{Barbieri}
M. Barbieri \emph{et al.}, Phys. Rev. Lett., {\bf 91}, 227901 (2003).
\bibitem{Bourennane}
 M. Bourennane \emph{et al.}, Phys. Rev. Lett., {\bf 92}, 87902 (2004).
\bibitem{ZollerIonTrap}
J.I. Cirac, P. Zoller, Phys. Rev. Lett., {\bf 74}, 4091 (1995).
\bibitem{Haeffner}
H. H\"affner \emph{et al.}, Nature, {\bf 438}, 643 (2005).
\bibitem{DetectEffKwiat}
P. G. Kwiat, A.M. Steinberg, R.Y. Chiao, P.H. Eberhard, M.D. Petroff, Phys. Rev. A, {\bf 48}, R867 (1993).
\bibitem{Guehne2}
O. G\"uhne \emph{et al.}, Phys. Rev. A, {\bf 66}, 062305 (2002).

\end{thebibliography}
\end{document}